\documentclass{aa}

\usepackage{graphicx}
\usepackage{txfonts}
\usepackage{booktabs}
\usepackage{multirow}
\usepackage{float}
\usepackage{placeins}
\usepackage{balance}
\usepackage{cuted}

\usepackage[usenames,dvipsnames]{xcolor}

\newcommand{\zssb}{z_\text{SSB}}

\newcommand{\corr}{}
\usepackage{amstext}

\begin{document} 

   \title{Distinguishing screening mechanisms with environment-dependent velocity statistics}
   \titlerunning{Distinguishing screening mechanisms with velocity statistics}
 
   \author{Magnus Fagernes Ivarsen
          \inst{1}
          \and
          Philip Bull
          \inst{2,3}
          \and
          Claudio Llinares
          \inst{4}
          \and
          David Mota
          \inst{1}
          }

   \institute{
     Institute of Theoretical Astrophysics, University of Oslo, P.O. Box 1029 Blindern, N-0315 Oslo, Norway
    \and
     California Institute of Technology, Pasadena, CA 91125, USA
    \and
         Jet Propulsion Laboratory, California Institute of Technology, 4800 Oak Grove Drive, Pasadena, California, USA
    \and
     Institute for Computational Cosmology, Department of Physics, Durham University, Durham DH1 3LE, UK
     }

   \date{\today}
  \abstract
   {Alternative theories of gravity typically invoke an environment-dependent screening mechanism to allow phenomenologically interesting deviations from general relativity (GR) to manifest on larger scales, while reducing to GR on small scales. The observation of the transition from screened to unscreened behavior would be compelling evidence for beyond-GR physics.}
   {We show that pairwise peculiar velocity statistics, in particular the relative radial velocity dispersion, $\sigma_\parallel$ , can be used to observe this transition when they are binned by some measure of halo environment.}
   {We established this by measuring the radial velocity dispersion between pairs of halos in N-body simulations for three $f(R)$ gravity and four symmetron models. We developed an estimator involving only line-of-sight velocities to show that this quantity is observable, and binned the results in halo mass, ambient density, and the isolatedness of halos.}
   {Ambient density is found to be the most relevant measure of environment; it is distinct from isolatedness, and correlates well with theoretical expectations for the symmetron model. By
binning $\sigma_\parallel$ in ambient density, we find a strong environment-dependent signature for the symmetron models, with the velocities showing a clear transition from GR to non-GR behavior. No such transition is observed for $f(R)$, as the relevant scales are deep in the unscreened regime.}
   {Observations of the relative radial velocity dispersion in forthcoming peculiar velocity surveys, if binned appropriately by environment, therefore offer a valuable way of detecting the screening signature of modified gravity.}

   \maketitle

\section{Introduction}

Observations in the solar neighborhood, on Earth, in binary pulsar systems, and now of gravitational waves \citep{2016arXiv160203841T} constrain possible deviations from general relativity (GR) with extraordinary precision \citep{2014LRR....17....4W}. This presents a formidable challenge to alternative theories of gravity, such as those being developed as a potential explanation of cosmic acceleration \citep{clifton_modified_2012, 2015PhR...568....1J, 2015arXiv151205356B};  the slightest disagreement with GR on the relevant scales can invalidate a given theory instantly.

With the partial exception of gravitational waves, which propagate over cosmological distances, the subjects of all of these tests are astrophysical systems that reside in dense galactic environments. Here, conditions are far from the cosmological average: the matter density, curvature, and gravitational potential may differ from the background by several orders of magnitude \citep{2015ApJ...802...63B}. One way of evading the stringent small-scale constraints while still allowing non-trivial deviations from GR on large scales is therefore to screen modifications to GR in certain environments. Several such screening mechanisms have been proposed, but all fall into one of three categories: Chameleon, K-mouflage, or Vainshtein screening \citep{brax_screening_2013, brax_distinguishing_2015}. In each case, deviations from GR are suppressed according to either the value of the scalar field or its derivatives, which vary depending on environmental factors such as ambient matter density or proximity to massive objects.

Screening mechanisms are a relatively generic prediction of viable modified gravity (MG) theories \citep{2012PhRvD..86d4015B}, and so detecting one at work would be a powerful indicator of beyond-GR physics. Observational tests of screening necessarily focus on the transition between the fully screened and unscreened regimes, where differences from GR phenomenology are expected to be most pronounced. For most viable cosmological theories, this transition is expected to occur at the outskirts of dark matter halos, where the matter density and gravitational potential begin to to approach their background values. A number of studies have looked for signatures of modified gravity in this region, for example through differences in halo or galaxy clustering statistics {\corr \citep[e.g.,][]{2005PhRvD..71f4030S,davis, 2008PhRvD..78l3524O, 2009PhRvD..79h3518S, 2013JCAP...10..027B, 2013PhRvD..88h4029W, 2013MNRAS.428..743L, 2014ApJS..211...23Z, clifton,2015PhRvL.114y1101L},} or the density, velocity, and potential profiles of individual halos \citep[e.g.,][]{2009arXiv0911.1829M, 2012PhRvD..85j2001L, 2013MNRAS.431..749C, 2013JCAP...10..012H, gronke_gravitational_2014, gronke_halos_2015, 2016arXiv160300056S}.

In this paper, we study the effect of modifications to GR on pairwise peculiar velocity statistics \citep{davis_integration_1977}, a relatively model-independent class of observables \citep{ferreira_streaming_1999} that probe scales that are well-matched to the expected screening transition. Some of these statistics were recently found to exhibit strong deviations from their GR behavior on quasi-linear and non-linear scales in simulations of $f(R)$ and Galileon theories \citep{hellwing_clear_2014}.

Here, we use similar simulations of $f(R)$ and symmetron theories to investigate the {\it \textup{environmental dependence}} of the deviations. We find that a clear, theory-dependent signature of screening can be observed when the ambient matter density, smoothed on scales on the order of the inter-halo separation, is chosen as the proxy for environment. Other proxies also have interesting theory-dependent behaviors, but are harder to interpret. We also clarify the relationship between theoretical pairwise velocity statistics and observable (line-of-sight) quantities, and propose a new estimator for the pairwise velocity dispersion $\sigma_\parallel$ (sometimes called $\sigma_{12}$). The result of our study is an observable that is sensitive to the theory-dependent properties of screening mechanisms, and a set of predictions for what those signatures should look like in $f(R)$ and symmetron gravity.

The paper is organized as follows. In Sect.~\ref{mg} we give an overview of the MG theories and simulations that we used, and outline predictions for screening effects based on previous work. In Sect.~\ref{velstat} we define several relevant velocity statistics, and clarify their relationship with observables. Then, in Sect.~\ref{envir}, we define two ways of measuring halo environment and briefly compare them. Finally, we present results from the simulations in Sect.~\ref{results}, and we conclude in Sect.~\ref{conclusions}.

\section{Modified gravity simulations} \label{mg}

In this section, we describe the suite of N-body simulations used in this study. In addition to a reference GR/$\Lambda$CDM simulation, we used three simulations of $f(R)$ modified-gravity models \citep{hu_models_2007} and four simulations of symmetron modified-gravity models \citep{hinterbichler_symmetron_2011}. Both types of MG model screen the fifth force in high-density environments by way of a chameleon-like screening that depends on the value of an additional scalar field. In what follows, we provide a brief overview of the particular modified-gravity models that were used, and describe the basic properties of the simulations. More details about the simulations can be found in \citet{llinares_isis:_2014}.

\subsection{$f(R)$ gravity}

A simple way of modifying GR is to replace the Ricci scalar, $R$, in the Einstein-Hilbert action with a more general function $f(R)$,
\newcommand{\mpl}{M_\text{pl}}
\begin{equation}
S = \frac{\mpl}{2}\int \! \mathrm{d}^4x\sqrt{-g}f(R) + S_m(g_{\mu\nu},\Psi),
\end{equation}
where $\mpl=(8\pi G)^{-1/2}$, $G$ is the bare gravitational constant, $g$ is the determinant of the spacetime metric $g_{\mu\nu}$, and $S_m$ is the action for matter fields $\Psi$. By applying an appropriate conformal transformation, these models can be written as an effective scalar field theory for any choice of the function $f(R)$ \citep{clifton_modified_2012}. The resulting scalar field, $f_R\equiv\mathrm{d}f/\mathrm{d}R$, gives rise to a so-called fifth force,
\begin{equation}
F_\phi = - \frac{1}{2}\nabla f_R.
\end{equation}
Screening of the fifth force (e.g., to satisfy solar system constraints) is achieved using the chameleon mechanism. The scalar field $f_R$ propagates in an effective potential,
\newcommand{\Mpl}{M_\text{Pl}}
\begin{equation}
V(f_R) = \frac{\Mpl^2}{2}\frac{Rf_R-f}{f_R^2},
\end{equation}
with a density-dependent effective mass \citep{amendola_dark_2010}. 
In high-density environments, the mass becomes high and the effective potential steepens, stopping the field from propagating freely, and sending $F_\phi \to 0$.

Models are specified by choosing a particular form for $f(R)$. A commonly used parametrization introduces a single extra parameter $f_{R0}$, the value of the scalar field at the present epoch \citep{hu_models_2007}. This parameter controls the degree to which the model deviates from GR, that is, $f_{R0} \to 0$ is the GR limit. The three $f(R)$ models for which we have simulations differ only by the field value: $f_{R0} = \{ 10^{-6}, 10^{-5}, 10^{-4} \}$ (in order of weakest to strongest deviation from GR). We note that these models are disfavored by current data \citep[e.g.,][]{2013ApJ...779...39J, 2015arXiv150201590P}, but nevertheless remain useful for the purposes of illustration.

\begin{table}
\centering
\begin{tabular}{@{}llll@{}}
\toprule
Name & $\beta$ & $\zssb$ & $L$ \\ \midrule
Weaker coupling (\texttt{symmA})   & 1       & 1       & 1   \\
Stronger coupling (\texttt{symmC}) & 2       & 1       & 1   \\
Early SSB (\texttt{symmB})       & 1       & 2       & 1   \\
Super-early SSB (\texttt{symmD}) & 1       & 3       & 1   \\ \bottomrule
\rule{0pt}{4ex}    
\end{tabular}

\begin{tabular}{@{}lll@{}}
\toprule
Name & $f_{R0}$  & $n$ \\ \midrule
Weak (\texttt{fr6})            & $10^{-6}$ & 1   \\
Medium (\texttt{fr5})          & $10^{-5}$ & 1   \\
Strong (\texttt{fr4})          & $10^{-4}$ & 1   \\ \bottomrule
\rule{0pt}{4ex}    
\end{tabular}
\centering
\caption{Parameters of the four realizations of symmetron gravity and three realizations of $f(R)$ gravity in the N-body simulations of \citet{llinares_isis:_2014}. The names of the simulations from the original paper are given in parentheses.}
\label{flavours}
\end{table}

\subsection{Symmetron gravity}

Symmetron gravity explicitly introduces a new scalar field into the gravitational action \citep{2005PhRvD..72d3535P, hinterbichler_symmetron_2010, hinterbichler_symmetron_2011},
\begin{equation} \label{symm}
S = \int \! \mathrm{d}^4x\sqrt{-g}\left[ \frac{\mpl}{2}R - \frac{1}{2}g^{a b}\partial_a \phi \partial_b \phi - V(\phi)\right] + S_m(\tilde{g}_{\mu\nu},\Psi),
\end{equation}
which means that the Einstein-Hilbert term has been supplemented by a kinetic term and potential $V$ for the scalar field $\phi$. A key element of symmetron theories is the choice of a symmetry breaking form for the effective potential, such as \citep{hinterbichler_symmetron_2011}
\begin{equation} \label{eq:Vsymm}
V_\text{eff}(\phi) = \frac{1}{2}\left( \frac{\rho}{M^2}-\mu^2\right)\phi^2+\frac{1}{4}\lambda \phi^4,
\end{equation}
where $\rho$ is the ambient matter density. The mass $M$ is given by
\begin{equation}
M^2 = 2\Omega_{{\rm M}0}\,\rho_{c0}\,L^2(1+\zssb)^3,
\end{equation}
with free parameters $L$, the effective range of the fifth force, and $z_\text{SSB}$, the redshift at which the symmetry spontaneously breaks (SSB). A third free parameter, the coupling strength $\beta$, is related to the vacuum expectation value of the field ($\beta~=~\phi_0 M_\text{pl}/M^2$, where $\phi_0~=~\mu/\sqrt{\lambda}$). The fifth force in these models is given by
\begin{equation}
F_\phi = \frac{\phi\nabla \phi}{M^2}.
\end{equation}
Symmetron gravity is distinguished by having an effective potential for which the symmetry is broken only in regions of low ambient density; when $\rho \ll M^2 \mu^2$, the reflection symmetry $\phi \rightarrow -\phi$ is spontaneously broken, and the field acquires the vacuum expectation value $|\phi|~=~\phi_0$. In regions of high ambient density, the symmetry remains unbroken, and the fifth force is screened.

The simulations that we used all assume the same range for the fifth force, $L$, but differ in coupling strength and SSB redshift. Two simulations have the same $z_{\rm SSB} = 1$, but differ in coupling strength $\beta$ by a factor of 2 (weaker vs. stronger coupling), while the other two keep the weaker coupling strength, but have earlier SSB redshifts of $z_{\rm SSB} = 2$ and $3,$ respectively. These model parameters are summarized in Table~\ref{flavours}.

\subsection{N-body simulation parameters} \label{simulations}

We used eight simulations in total: one for GR/$\Lambda$CDM, three for $f(R)$ gravity, and four for symmetron gravity, as previously described in \citet{llinares_isis:_2014}, all with identical initial conditions and cosmological expansion histories. The simulations were made with the N-body code ISIS \citep{llinares_isis:_2014}, which is based on the \texttt{RAMSES} {\corr adaptive mesh refinement} code \citep{2010ascl.soft11007T}. We used boxes of side length 256 $h^{-1}$Mpc and 512$^3$ dark matter particles per box, each of mass $9.26\times 10^9 \, h^{-1} M_\odot$. {\corr Seven levels of refinement were applied to a domain-uniform grid of 512 nodes per dimension, giving an effective resolution of 32,768 nodes per dimension at the deepest refinement level (which corresponds to a maximum spatial resolution of $7.8\, h^{-1}{\rm kpc}$).} Snapshots were output at $z=0$ for each simulation, and halos were identified using the \texttt{ROCKSTAR} code \citep{behroozi_rockstar_2013}. Halos with fewer than 50 particles (corresponding to $M_{\rm halo} < 5\times 10^{11} \, h^{-1} M_\odot$) were considered unresolved, and discarded from the halo catalogs. We also neglected light-cone effects, as the simulation boxes are sufficiently small. {\corr We note that all our calculations here use the dark matter halos and not the particles from the simulation.}

\section{Peculiar velocity statistics} \label{velstat}

Pairs of galaxies or dark matter halos have a mean tendency to fall toward one other, meaning that their mean {\it \textup{relative}} velocity is non-zero. This observation has been used to construct a class of pairwise velocity statistics, first introduced by \citet{davis_integration_1977}, and with subsequent development by {\corr \citet{gorski_pattern_1988}, \citet{consortium_evolution_1998}, \citet{ferreira_streaming_1999}, \citet{juszkiewicz_dynamics_1999}, \citet{2000Sci...287..109J}, \citet{2001MNRAS.322..901S}, \citet{2004PhRvD..70h3007S}, \citet{2015A&A...583A..52M},} and others. As their basic building block, these use the relative radial velocity between the halos, that is, the component along the separation vector between them, {\corr which effectively picks out a preferred direction.}

In this paper, we consider two pairwise statistics: $v_\parallel(r)$ (sometimes called $v_{12}(r)$), the mean relative radial velocity as a function of the pair separation, $r$; and $\sigma^2_\parallel(r)$, the variance (dispersion) of $v_\parallel$. Variants of $v_\parallel$, for instance, with a mass weighting, have recently been used to detect the coherent velocity signal from the kinetic Sunyaev-Zel'dovich effect \citep{2012PhRvL.109d1101H} {\corr and search for signatures of modified gravity in the linear growth rate \citep{2015A&A...583A..52M}. They were also used to confirm the low matter density of the Universe \citep{2000Sci...287..109J}}. More importantly for our uses, a variant of $\sigma^2_\parallel$ was recently demonstrated to be particularly sensitive to modified gravitational dynamics in simulations \citep{hellwing_dynamics_2014} \citep[see also][]{2009ApJ...695L.145L}, and we therefore focus on this quantity in what follows. Notably, $\sigma_\parallel^2$ includes some non-linear terms that average out in $v_\parallel$ \citep{2004PhRvD..70h3007S}, but which are relevant on scales that are strongly affected by (e.g.) $f(R)$ gravity \citep{2013MNRAS.428..743L}.

\begin{figure}
\centering
  \includegraphics[width=0.43\textwidth]{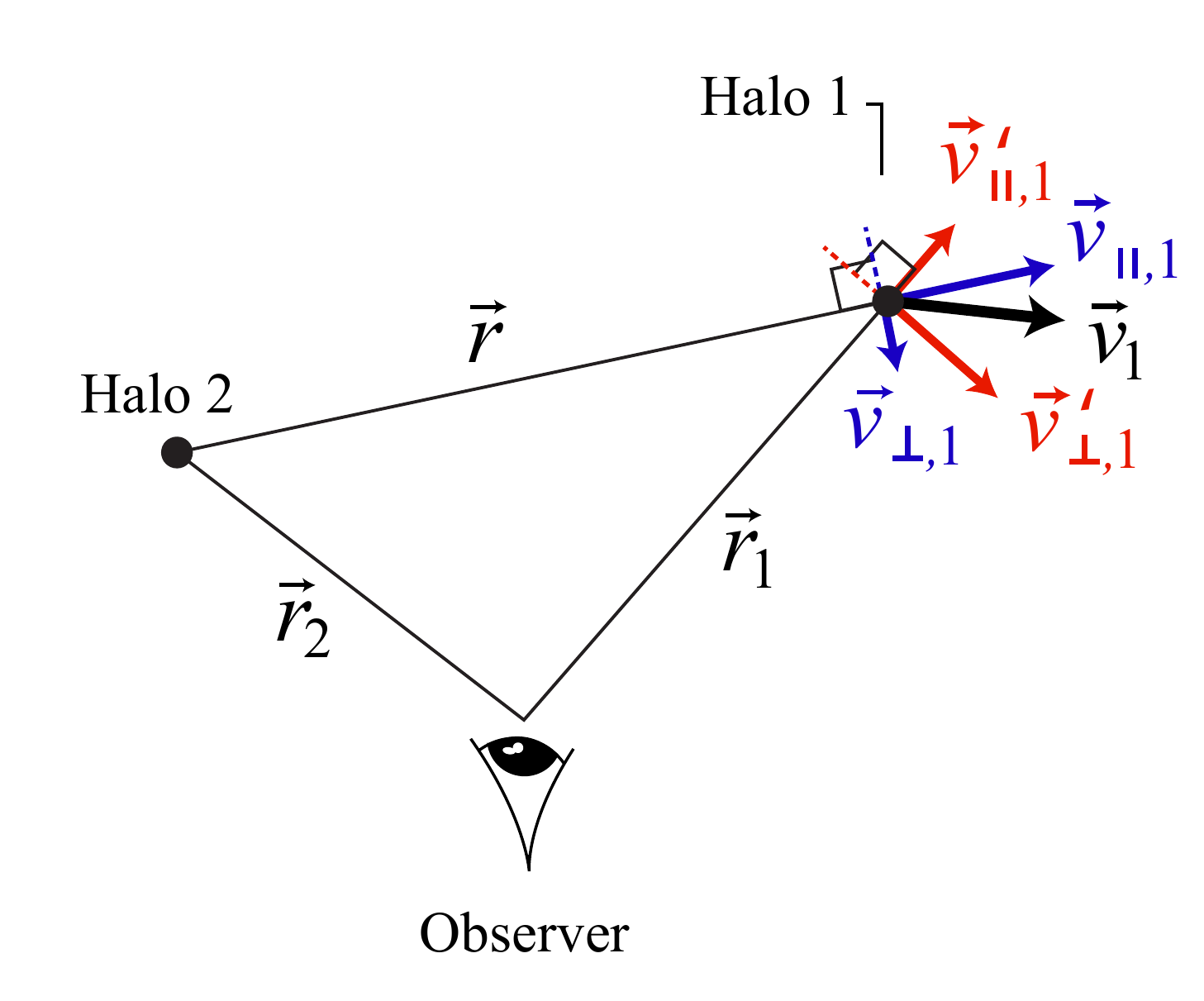}
  \caption{Geometry of halo pairs with respect to an observer. Velocities parallel or transverse to the line of sight ($v_\parallel^\prime, \vec{v}_\perp^\prime$) are shown in red, while velocities parallel
or transverse to the connecting vector $\vec{r}$ between the halos ($v_\parallel, \vec{v}_\perp$) are shown in blue.}
  \label{diagram}
\end{figure}

We begin by defining some geometrical quantities (Fig.~\ref{diagram}). The separation vector between halos is $\vec{r} \equiv \vec{r}_i - \vec{r}_j$, where $\vec{r}_{i,j}$ are the position vectors of halos $i,j$ from the observer. The relative velocity, $\vec{v}_i - \vec{v}_j$, can be decomposed into 1+2 components: $v_\parallel$, the radial component, along the connecting vector $\vec{r}$; and $\vec{v}_\perp$, a 2D vector in the space transverse to $\vec{r}$. Likewise, the velocity of each halo can be decomposed into components parallel ($s_i \equiv \vec{v}_i \cdot \hat{\vec{r}}_i$) and perpendicular ($\vec{t}_i$) to the line of sight (hats denote unit vectors). The magnitude of the halo velocity is then defined through $v_i^2 = {s_i^2 + |\vec{t}_i|^2}$.

Typically, only the line-of-sight (LOS) component of the velocity can be measured observationally, but $v_\parallel$ clearly depends on components both parallel and transverse to the LOS for a general pair geometry. As such, some form of deprojection method is needed to recover estimates of $v_\parallel$ and $\sigma^2_\parallel$ from data. \cite{ferreira_streaming_1999} achieved this by deriving an estimator for the mean radial velocity purely in terms of the observable LOS velocities and angles,
\begin{equation}\label{v12-est}
 \tilde{v}_\parallel(r) = \frac{\sum_{i,j} (s_i-s_j)\cos \theta_{ij}}{\sum_{i,j} \cos^2\theta_{ij}}.
\end{equation}
A distant-observer approximation has been used, such that $\cos \theta_{ij} \equiv \hat{\vec{r}}\cdot \hat{\vec{l}}$, where $\hat{\vec{l}} \equiv \hat{\vec{r}}_i \approx \hat{\vec{r}}_j$. Note that \cite{ferreira_streaming_1999} used a slightly different definition, $\hat{\vec{l}} \equiv (\hat{\vec{r}}_i + \hat{\vec{r}}_j)/2$, which is the mean of the LOS directions. Summations are performed over all pairs within a given separation bin, $r_n \le r < r_{n+1}$.

We are unaware of an equivalent estimator for the dispersion, and so derive one in Appendix \ref{app:estimators}, obtaining
\begin{equation}
\tilde{\sigma}^2_\parallel(r) \approx \frac{\sum_{i,j} \cos^2 \theta_{ij} (s_i - s_j)^2}{\sum_{i,j} \cos^4 \theta_{ij}} \label{eq:sigpar}
\end{equation}
in the limit of small $\theta_{ij}$. More general expressions, including an estimator for the transverse dispersion, $\tilde{\sigma}^2_\perp(r)$, are given in the appendix.

We note that $\tilde{\sigma}^2_\parallel(r)$ is different from the quantity $\sigma^2_{12}(=\!\!\sigma^2_{\rm los})$ used by \cite{hellwing_dynamics_2014}. The latter was defined by the \cite{consortium_evolution_1998} as a quantity that is ``somewhat closer to measurements accessible in galaxy redshift surveys'' than the actual radial dispersion, ${\sigma}^2_\parallel(r)$. In our notation,
\begin{equation}
\sigma_{\rm los}^ 2(r_p)\equiv\frac{\int\limits_{-\infty}^\infty \! \xi(r)\, \left [\frac{1}{2}\sin^2\theta_{ij}\,\sigma_\perp^2 + \cos^2\theta_{ij}\,(\sigma_\parallel^2-\langle v_\parallel \rangle^2) \right ] \; dl}{\int\limits_{-\infty}^\infty \! \xi(r)\; dl }, \label{eq:sig12}
\end{equation}
where $r_p \equiv r \sin \theta_{ij} = \sqrt{r^2 - l^2}$ is the projected pair separation, $\xi(r)$ is the halo correlation function, and the integration is along the LOS direction. The term in square brackets is the (centered) projected velocity dispersion, equivalent to the LOS velocity dispersion for a given pair geometry $(r,l)$. In our notation, $\sigma^2_\perp$ is the centered variance of $\vec{v}_\perp$. The factor of $\frac{1}{2}$ arises because one component of $\vec{v}_\perp$ is orthogonal to the LOS and therefore does not contribute to the LOS dispersion. The integration over the LOS in Eq.~(\ref{eq:sig12}) is equivalent to an ensemble average over all pairs with the same $r_p$, where $\xi$ acts as a weight factor (normalized in the denominator) that specifies the fraction of pairs with separation $r$. In the ensemble, we therefore have
\begin{equation}
\sigma_{\rm los}^ 2(r_p) \equiv \sum_{i,j} (s_i - s_j)^2 - \left ( \sum_{i,j} (s_i - s_j) \right )^2,
\end{equation}
where the sums are over all pairs with fixed $r_p$. While this {\it \textup{is}} an observable, as claimed by \cite{consortium_evolution_1998}, it is not a direct estimate of the variance of $v_\parallel$ as our estimator $\tilde{\sigma}^2_\parallel$ is, and we therefore
did not use it here.

\begin{figure}[t]
\centering
\includegraphics[width=0.49\textwidth]{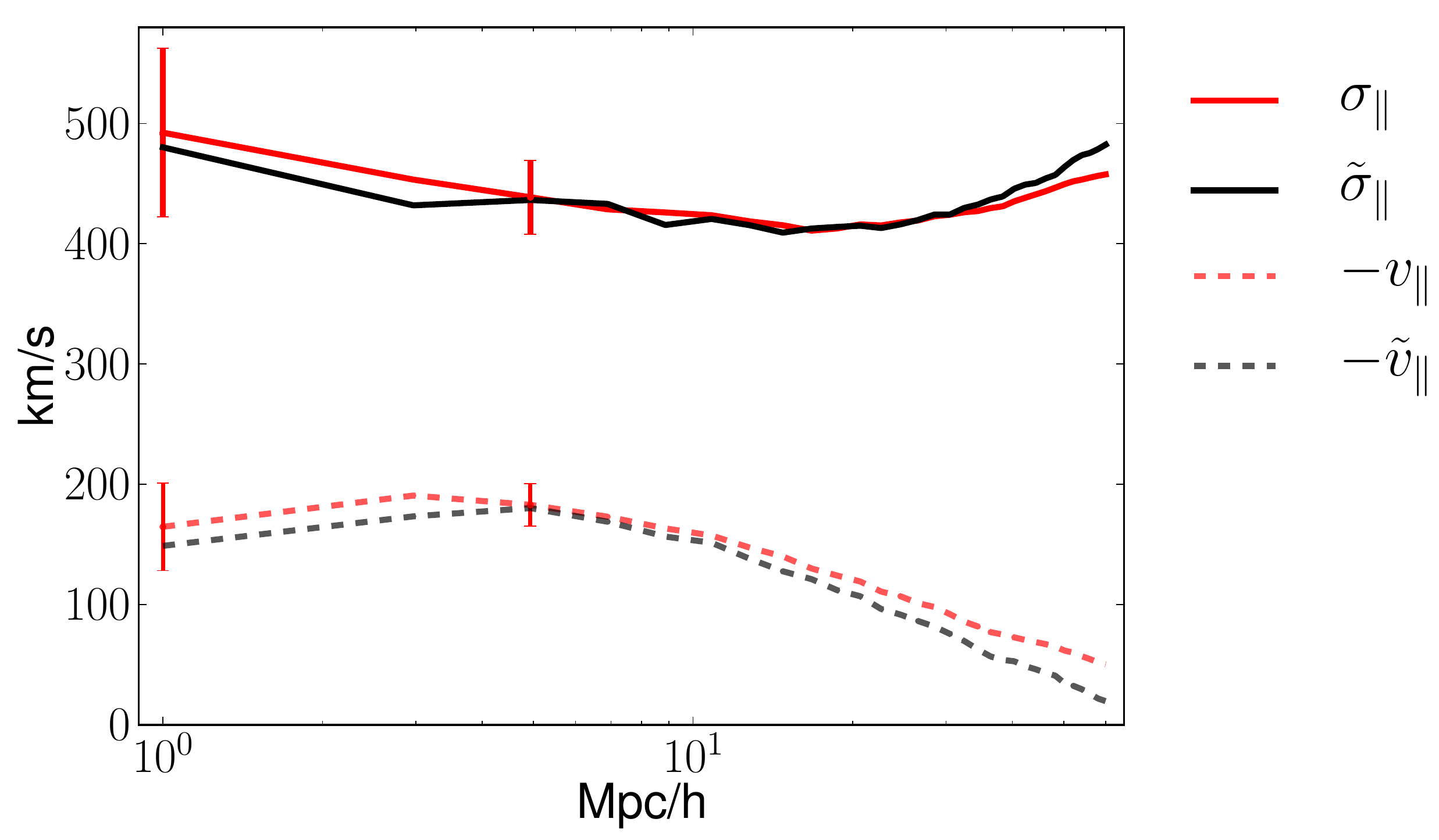}
\caption{Comparison of the mean relative radial velocity, $v_\parallel(r)$, and the square root of its variance, $\sigma_\parallel(r)$, measured from $\Lambda$CDM simulation halos using full 3D information (red lines), and LOS-velocity estimators only (black lines). Quantities are shown in bins of pair separation, $r$. {\corr Bootstrap errors (68\% CL) are shown for the $r=(1, 5)\, h^{-1}{\rm Mpc}$ bins; we note that they are significantly larger than the differences between the black and red curves.}}
\label{fig:sigparest}
\end{figure}

In the remainder of the paper, we use the true radial velocity dispersion $\sigma^2_\parallel$, calculated from the full 3D velocity information in the simulated halo catalogs. This can then be connected to observations using the estimator $\tilde{\sigma}^2_\parallel$ from Eq.~(\ref{eq:sigpar}). In Fig.~\ref{fig:sigparest} we show a comparison between the two quantities for the $\Lambda$CDM simulation; the agreement is good, although $\tilde{\sigma}^2_\parallel$ is naturally noisier.

\section{Measures of environment and screening}
\label{envir}

As discussed in Sect.~\ref{mg}, many modified gravity theories exhibit an environment-dependent screening of the fifth force. In this section, we describe a few possible ways of defining the environment of a dark matter halo, each of which is expected to trace the degree of screening in some way. These measures are then used to divide the simulated halo catalogs into multiple environment-dependent samples in Sect.~\ref{results}.

\begin{table}
\centering
\begin{tabular}{@{}ll|c@{}}
\toprule
Model & & $\mu_{200} \,[ h^{-1} M_\odot ]$ \\
\midrule
\multirow{3}{*}{$f(R)$} & Weak   & $1.0 \times 10^{13}$         \\
 & Medium & $1.0 \times 10^{14}$         \\
 & Strong & $3.0 \times 10^{16}$ \\
\midrule
\multirow{4}{*}{Symmetron} & Weaker coupling   & $4.5\times 10^{11}$ \\
 & Stronger coupling & $4.5\times 10^{11}$ \\
 & Early SSB       & $2.8\times 10^{12}$ \\
 & Super-early SSB & $2.0\times 10^{13} $ \\
\bottomrule
\rule{0pt}{4ex}    
\end{tabular}
\caption{Rescaling mass (Eqs.~\ref{rescf} and \ref{rescs}) for the simulated MG theories listed in Table \ref{flavours}. Halos with mass $M_{\rm 200} \gg \mu_{200}$ are expected to be fully screened, while those with $M_{\rm 200} \simeq \mu_{200}$ will be partially screened.}
\label{rescmass}
\vspace{-1em}
\end{table}

\subsection{Halo rescaling mass}

As shown in \citet{gronke_universal_2015}, the halo mass itself is a fairly universal predictor of whether a halo is screened. Using an empirical fit of the fifth-force profiles of simulated halos with NFW density profiles, they defined the rescaling mass, $\mu_{200}$ , which is a characteristic mass scale above which the fifth force is expected to be screened. For a given theory, $\mu_{200}$ is defined as the halo mass for which the mass-weighted mean of the fifth force in the halo is half its maximum (unscreened) strength. Partial screening is expected in a band of halo masses around $\mu_{200}$. For the symmetron and $f(R)$ models considered in this paper, \citet{gronke_universal_2015} found the fitting functions
\begin{align}
\mu_{200}^\text{Symm} &\approx 2\times 10^{10}\, h^{-1}M_\odot \left(\frac{L}{h^{-1}\text{Mpc}}\right)^3(1+\zssb)^{4.5}\; \label{rescf} \\
\mu_{200}^{f(R)} &\approx 10^{13}\, h^{-1}M_\odot \left(\frac{f_{R0}}{10^{-6}}\right)^{1.5}. \label{rescs}
\end{align}
While these calculations are based on the strength of the fifth force affecting a test mass around an individual halo, the forces between nearby halos with similar masses can be expected to follow a similar pattern. This, in turn, will affect their clustering properties, so that we should expect clustering or velocity statistics for halos in mass bins well above the rescaling mass to be screened (i.e., reduced to their GR behavior). Conversely, halos in mass bins around or far below the rescaling mass will be partially or fully unscreened, respectively, so that deviations from GR should become apparent in the clustering statistics.

The values of the rescaling mass for each of the models considered here are given in Table \ref{rescmass}. The $f(R)$ models, even the weakest with $f_{R0}=10^{-6}$ ,  have high rescaling masses, $\mu_{200} \ge 10^{13}\, h^{-1} M_\odot$, which suggests that most halos from the simulations (which have typically much lower masses) should be partially or fully unscreened. The rescaling masses are substantially lower for all of the symmetron models except for super-early SSB, however, and a range of screening states are therefore expected for these models.

\begin{table*}[t]
\centering
\begin{tabular}{@{}l|lllll@{}}
\toprule
\multicolumn{6}{c}{Halo mass bins, $\{ M_j \}$}          \\ \midrule
Bin edges in $M_{\rm halo}$ [$h^{-1} M_\odot$] & $[5\times 10^{11}, 10^{12} )$ & $[ 10^{12}, 2.2\times 10^{12})$ & $[2.2\times 10^{12}, 5\times 10^{12})$ & $[5\times 10^{12}, 10^{13})$ & $[10^{13},\infty )$ \\
$N_{\rm halos}$ & $2.9\times 10^4$ & $2.7\times 10^4$ & $1.7\times 10^4$ & $7.9\times 10^3$ & $7.4\times 10^3$ \\
$\langle {M}_{\rm halo}\rangle$  [$h^{-1} M_\odot$] & $7.1\times 10^{11}$ & $ 1.4\times 10^{12}$ & $3.2\times 10^{12}$ & $6.8\times 10^{12}$ & $2.3\times 10^{13}$ \\
$\langle b \rangle_{\rm halo}$ & 0.93                & 0.99                 & 1.10                & 1.23                & 1.65                \\
\bottomrule
\end{tabular}
\newline
\vspace*{0.4cm}
\newline
\begin{tabular}{@{}l|llllll@{}}
\toprule
\multicolumn{7}{c}{Ambient density bins, $\{ \delta_j \}$}  \\ \midrule
Bin edges in $\delta$     & $[-1.00,-0.70)$     & $[-0.70,-0.35)$      & $[-0.35,0.27)$      & $[0.27,1.99)$       & $[1.99,10.00)$      & $[10.00,\infty)$    \\
$N_{\rm halos}$ & $2.7\times 10^4$ & $1.3\times 10^4$ & $1.1\times 10^4$ & $1.2\times 10^4$ & $1.2\times 10^4$ & $1.3\times 10^4$ \\
$\langle {M}_{\rm halo}\rangle$ [$h^{-1} M_\odot$] & $9.6\times 10^{11}$ & $ 1.3\times 10^{12}$ & $1.6\times 10^{12}$ & $2.1\times 10^{12}$ & $2.9\times 10^{12}$ & $6.1\times 10^{12}$ \\
$\langle b \rangle_{\rm halo}$     & 0.97  & 1.01                & 1.04                 & 1.08                & 1.14                & 1.30                         \\
\bottomrule
\end{tabular}
\newline
\newline

\caption{Mean mass and bias for halos binned by mass (top) and ambient density (bottom), for the $\Lambda$CDM simulation. The high level of shot noise prevents a reliable measurement of bias from clustering in the simulation, therefore we estimated it by averaging over the halo model bias, $b(M_i)$, for the halos in each bin $j$, i.e., $\langle b\rangle^{(j)}_{\rm halo} = \sum_{i \,\in\, {\rm bin}\, j} b(M_i) / N^{(j)}_{\rm halos}$.
}
\label{bins}
\vspace{-1em}
\end{table*}

\subsection{Isolatedness of halos}\label{isolatedness}

In the simple halo model, the properties of dark matter halos depend only on their mass. Evidence from observations and simulations suggests that the environment in which a halo forms is also important, however; an environment-dependent assembly bias \citep{2007MNRAS.374.1303C} has been seen in the clustering of some observed galaxy samples, for example \citep{2015MNRAS.452.1958H}. Various definitions of environment have been used to probe such effects, but many of them essentially reduce to proxies for halo mass \citep{haas_disentangling_2012}. A more satisfactory definition of the environment of a halo is its isolatedness, which can be defined through \citep{haas_disentangling_2012}
\begin{equation} \label{isol}
  D_{N,f} \equiv \frac{d_{N}({M_N\geq f M})}{r^{\rm vir}_N}.
\end{equation}
Here, $D_{N,f}$ is the distance from the halo to the $N^\text{th}$ nearest neighbor whose mass exceeds $f$ times the mass of the halo, divided by the virial radius of the neighbor, $r^{\rm vir}_N$. Halos with $D_{N,f}$ greater than some threshold are considered to be isolated. This definition has the advantage of being insensitive to halo mass.

We follow \citet{zhao_2011} and \citet{winther_environment_2012} in defining halos with $D_{1,1}~>~10$ to be isolated, which are those farther away from the nearest more-massive halo than ten times its virial radius. We also divide the halos into low-mass ($5\times 10^{11}\leq M_{\rm halo} \leq 2\times 10^{12} \, h^{-1} M_\odot$) and high-mass ($M_{\rm halo} > 2 \times 10^{12} \, h^{-1} M_\odot$) samples, resulting in a total of four bins in isolatedness (isolated vs. clustered) and mass (low vs. high). 

It has been suggested in the literature \citep[e.g.,][]{zhao_2011, winther_environment_2012, gronke_halos_2015} that isolatedness is a proxy for the ambient density around the halo. This is not supported by our results, however (see Fig.~\ref{haloos} and Sect.~\ref{envircompar}, below), nor specifically by the original \cite{haas_disentangling_2012} paper. Whether isolatedness is a useful measure of environment for studying modified gravity effects is also unclear: it has been found to correlate with chameleon-like screening by some authors \citep{zhao_2011,winther_environment_2012,falck_2015}, but has also been found not to be associated with signs of screening by others \citep{gronke_halos_2015}.

\subsection{Ambient density field} \label{sec:density}

As discussed in Sect.~\ref{mg}, chameleon-like screening mechanisms respond to the ambient matter density: the fifth force vanishes in regions where $\rho$ is sufficiently large. Binning halos according to their ambient density should therefore be a useful way of studying the screening behavior. The density distribution is scale-dependent, however, so that the mean density estimated over a finite region will depend on a smoothing scale related to the region size. The density that enters (e.g.) in the symmetron potential, Eq.~(\ref{eq:Vsymm}), is a local quantity, with no implied averaging over extended regions, therefore it is not clear what smoothing scale to use for estimating $\rho$. This is not a problem in simulations because local values for the potential and field value are defined down to the simulation resolution.

\begin{figure}
\vspace{0em}
\centering
\includegraphics[scale=0.39]{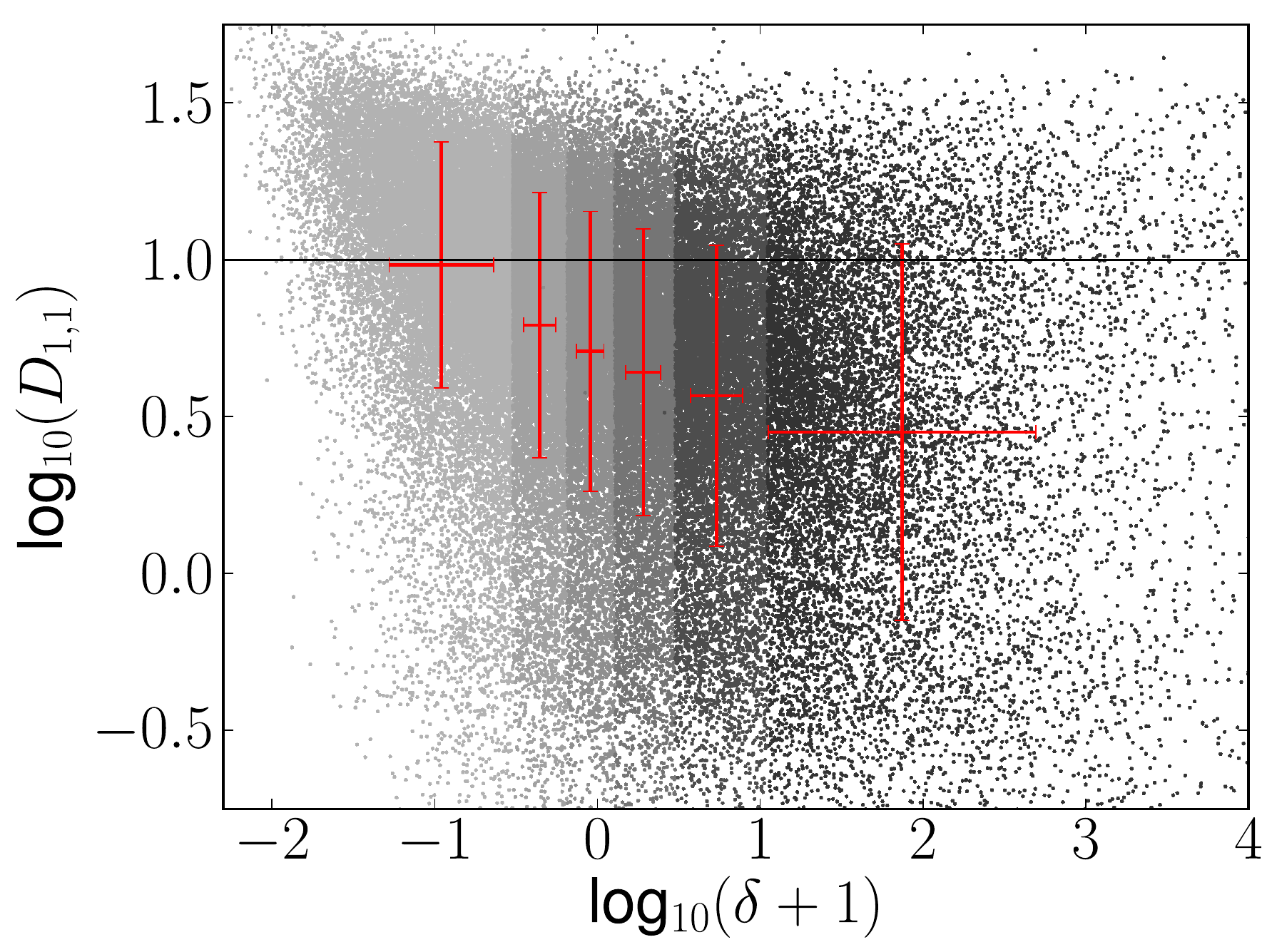}
\caption{Isolatedness vs. ambient density for halos of mass $M_{\rm halo}~\ge~5~\times~10^{11} \,h^{-1} M_\odot$, taken from the $\Lambda$CDM simulation. The horizontal line divides the isolated ($\log_{10} D_{1,1} \ge 1$) and clustered subsamples. The gray shading of the points denotes the binning in ambient density; red errorbars denote the mean and standard deviation of the ambient density and isolatedness in each ambient density bin.
}
\label{haloos}
\end{figure}

\begin{figure*}[t]
\centering
\includegraphics[scale=0.44]{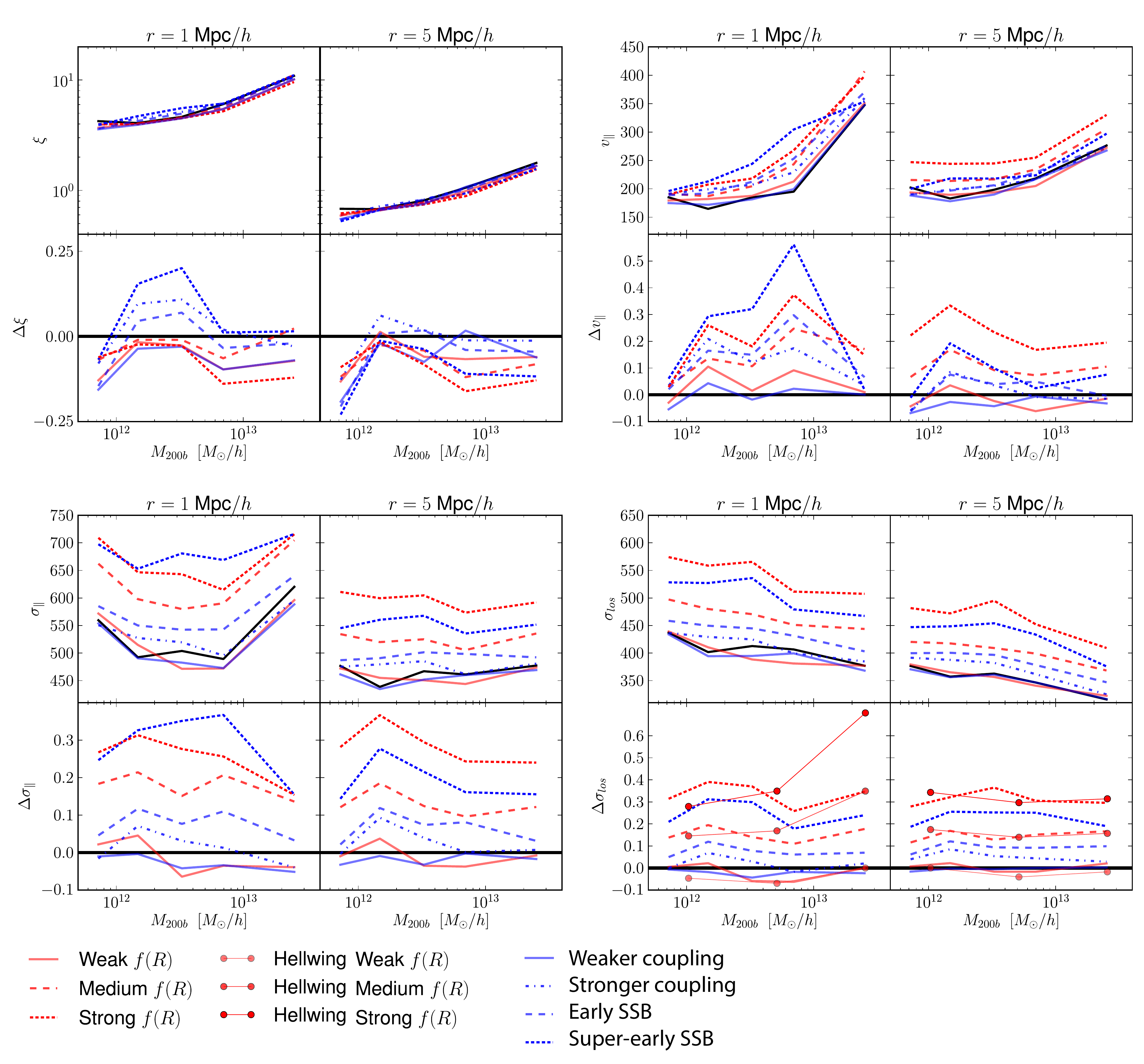}
\caption{Comparison of statistics across models: $\xi(r)$ (top left panel), $v_\parallel(r)$ (top right), $\sigma_\parallel(r)$ (bottom left), and $\sigma_{\rm los}(r)$ (bottom right). The left and right columns in each panel are for pair separations $r=1$ and $5$ $h^{-1}$Mpc, respectively. The upper rows show absolute values of the statistics, while the lower rows show the fractional deviation from GR ($\Delta_X = X_\text{MG}/X_\text{GR}-1$ for quantity $X$). The red points are taken from the $f(R)$ model results presented in \citet{hellwing_clear_2014}. {\corr (These
authors binned in $M_{200c}$, the mass measured at the radius where $\Delta=200$ with respect to {\it \textup{critical}} density; throughout this paper, we have consistently used $M_{200b}$, which is measured with respect to the background matter density. The difference in the curves when $M_{200c}$ is used is negligible, however.)}
}
\label{compar}
\vspace{-0.2em}
\end{figure*}

For our problem, a reasonable choice is to consider the density field smoothed on scales on the order of the inter-halo separation, since this is the smallest scale on which we can reliably construct an estimate of the density field from a halo catalog. This can be consistently estimated by performing a Delaunay triangulation \citep{delaunay_sur_1934} on the point distribution of halos, which is similar to a full Delaunay tessellation field estimation  \citep{schaap_continuous_2000}. Delaunay triangulation creates a set of tetrahedra that covers the volume spanned by the halo catalog, with each vertex coinciding with the position of a halo. The triangulation yields smaller tetrahedra around halos in regions with a high halo number density (i.e., many neighbors) and larger ones around more isolated halos. Dividing the halo mass by the volumes of the tetrahedra connected to that halo gives a point-wise measure of density \citep{Weygaert2009},
\begin{equation}\label{density}
 \rho_i = \frac{4\, M_{\text{halo},i} }{\sum\limits_j V_{i,j}},
\end{equation}
where the index $j$ labels the (connected) Delaunay tetrahedra, and the factor of 4 is a dimensional normalization constant, resulting from the fact that each tetrahedron is counted four times.
The sum $W_i=\sum_j V_{i,j}$ is called the contiguous Voronoi cell around point $i$. This definition of point-wise density ensures that integrating the density over the simulation space yields the total mass of the halos used.

The ambient density contrast $\delta_i$ at point $i$ is defined by $(1~+~\delta_i)~=~{\rho_i}/{\bar{\rho}}$, where $\bar{\rho}$ is the mean density. We use six bins in $\delta$ in our analysis, as summarized in Table~\ref{bins} and shown in Fig.~\ref{haloos}. We performed Delaunay triangulation by using a \texttt{MATLAB} routine, which in turn uses the Computational Geometry Algorithms Library, \texttt{CGAL} \citep{cgal:eb-15b}.

\subsection{Relationship between environment measures} \label{envircompar}

Isolatedness (Eq.~\ref{isol}) and ambient density (Eq.~\ref{density}) are expected to measure environment in substantially different ways. The former is constructed to be independent of halo mass, for example, while the latter will be correlated with mass, since high-mass halos tend to form in regions of higher density \citep{haas_disentangling_2012, grutzbauch_211,chen_2015}.

Figure~\ref{haloos} shows the isolatedness measure $D_{1,1}$ against ambient density for all halos in the $\Lambda$CDM simulation above a minimum mass cutoff. The correlation between the two measures is weak (correlation coefficient $r=0.03$), with each density bin containing halos of a wide range of isolatedness. Even the highest density bin ($\delta \ge 10$) contains a substantial fraction of isolated halos. Nevertheless, there is a slight tendency for isolated halos to form in regions of low density, which is
presumably due to the difficulty of forming more than one massive halo in voids.

This means that we expect some measures of environment to be more effective markers of screened regions or halos than others. Since the chameleon-like screening in symmetron models has an explicit dependence on the local density, we would expect the binning in ambient density $\delta_i$ to exhibit the clearest differences in observables between screened and unscreened regions (although this will also be correlated with mass-dependent effects). We compare the effectiveness of the different measures in distinguishing screened and unscreened models in the following section.


\section{Results} \label{results}

In this section, we investigate the dependence of the velocity statistics on halo mass, isolatedness, and ambient density for each of the simulated modified gravity models. We mostly focus on the quantity $\Delta \sigma_\parallel$ (see Sect.~\ref{velstat}), which is the fractional deviation of the radial velocity dispersion from its GR+$\Lambda$CDM value (where we define $\Delta_X = X_\text{MG}/X_\text{GR}-1$ for quantity $X$).

\subsection{Dependence on halo mass}\label{sec:mass}

In Fig.~\ref{compar} we show the absolute value and fractional deviation from GR for four different statistics, binned by halo mass with the binning specified in Table~\ref{bins}. Each bin is shown for two different halo pair separations, $r=1$ and $5\, h^{-1}$Mpc. We focus on $r=5\, h^{-1}$Mpc below because the $1\, h^{-1}$Mpc bin is considerably noisier {\corr (cf. the bootstrap errors in Fig.~\ref{fig:sigparest})}. Note that the noise due to sample variance is correlated between models because the simulations share the same initial conditions. Nevertheless, we include both in the figures to facilitate comparison with equivalent results in \citet{hellwing_clear_2014}.

The upper left panel of Fig.~\ref{compar} shows the halo-halo correlation function, $\xi_{\rm hh}(r)$, estimated using the Landy-Szalay estimator \citep{landy_bias_1993}, with 7,000-30,000 halos per mass bin (Table \ref{bins}). There is a clear trend of increasing correlation with mass for all models, which is expected; more massive halos are more strongly biased. The deviation $\Delta \xi$ becomes larger and more negative as we go from the weak to the strong $f(R)$ model, indicating a suppression of clustering on these scales. A similar but more confused picture can be seen for the symmetron models. There is no clear evidence for a dependence of $\Delta\xi$ on mass, although there is a spike feature in the second mass bin for all models. This is most likely due to a large scatter in the neighboring mass bins, which we tentatively observed through a (50:50) subsampling of each bin. Clearer trends in $\xi$ as a function of mass are observed by \cite{hellwing_clear_2014} because they averaged over six different realizations of the initial fluctuations; they found no strong evidence for a mass dependence either.

The upper right panel of Fig.~\ref{compar} shows the mean relative radial velocity, $v_\parallel$, in the same separation and mass bins. The $\Delta v_\parallel$ curves are more clearly separated for the different MG theories, with stronger modifications consistently giving larger fractional deviations. The dispersion of $v_\parallel$ is shown in the lower left panel. This quantity was calculated from the full 3D information in the simulations, not the estimator $\tilde{\sigma}_\parallel$. The pattern is more coherent than for the previous two quantities, with weaker modifications (e.g., weak $f(R)$ and weaker coupling symmetron) exhibiting $\Delta \sigma_\parallel \approx 0$, and stronger modifications (e.g. strong $f(R)$) giving up to $\sim 30\%$ deviations from GR. Again, there is no clear dependence on mass, either of $\sigma_\parallel$ or $\Delta \sigma_\parallel$.

\begin{figure}[t]
\centering
\hspace{-2em}\includegraphics[width=0.38\textwidth]{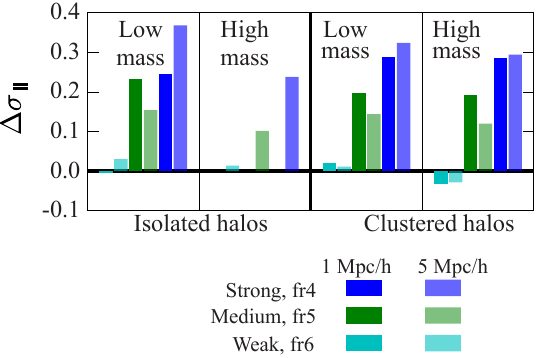}
\caption{Fractional deviation from GR ($\Delta \sigma_\parallel = \sigma_{\parallel,\text{MG}}/\sigma_{\parallel,\text{GR}}-1$) of the relative radial velocity dispersion, $\Delta\sigma_\parallel$, for the three $f(R)$ models, binned by isolatedness (isolated vs. clustered) and mass (low and high halo mass) as described in Sect.~\ref{isolatedness}. Pair separations $r=1$ and 5 $h^{-1}$Mpc are presented with bold and pale shading, respectively. The isolated high-mass halo bin was excluded for $r=1$ $h^{-1}$Mpc because
of the low number of halos within it.
}
\label{envirfig}
\end{figure}

\begin{figure}[t]
\centering
\hspace{-2em}\includegraphics[width=0.38\textwidth]{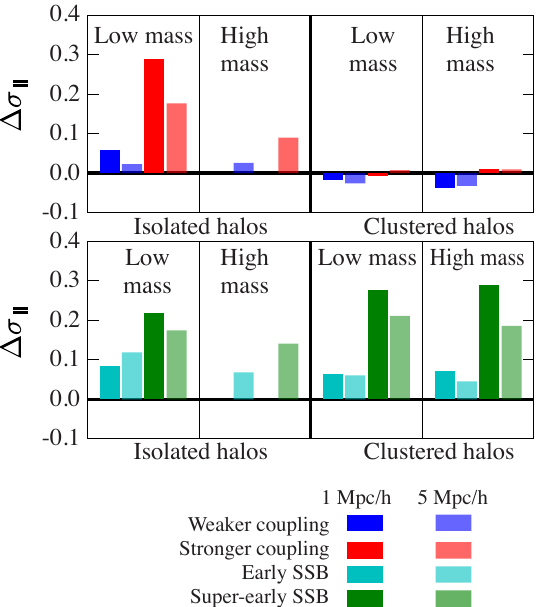}
\caption{Fractional deviation $\Delta\sigma_\parallel$ for the four symmetron models (see Fig.~\ref{envirfig} for key).
}
\label{envirfig2}
\end{figure}

Finally, the lower right panel of Fig.~\ref{compar} shows the dispersion of the relative LOS velocity, $\sigma_{\rm los}$. As for $\sigma_\parallel$, the trend with increasing strength of deviation from GR is clear, and the dependence on mass is weak (although $\sigma_{\rm los}$ can be seen to decrease slightly with increasing mass). We also find that $\Delta \sigma_\parallel \approx \Delta \sigma_{\rm los}$ in all cases, with similar scatter across the mass bins. We find excellent agreement with the results for $f(R)$ from \cite{hellwing_clear_2014} (red points in Fig.~\ref{compar}) at a separation of $r=5\, h^{-1}$Mpc, but less so at $r=1\, h^{-1}$Mpc, where they observed a much stronger mass dependence.

Overall, then, we find no clear evidence that the fractional deviations of the velocity statistics depend on halo mass. This is qualitatively similar to what was found in \cite{hellwing_clear_2014}, although some of the results differ in detail. The lack of a strong mass dependence can be understood in terms of the fifth-force profiles calculated by \cite{gronke_gravitational_2014} for the same symmetron and $f(R)$ theories considered here. At distances beyond the virial radius, $r \ge r_{\rm vir}$, the fifth force was found to be approximately constant as a function of mass in all models when $M_h \lesssim 10^{13} M_\odot$. (The fifth force decreases rapidly at higher $M_h$ for some models, e.g., weak $f(R)$, but is small in these models anyway, resulting in little overall change in velocity.)

\subsection{Dependence on isolatedness}

Figures~\ref{envirfig} and \ref{envirfig2} show the fractional deviations of the relative radial velocity dispersion, $\Delta \sigma_\parallel$, in bins of halo mass and isolatedness, $D_{1,1}$. As discussed in Sect.~\ref{isolatedness}, halos are considered isolated when the nearest more-massive halo is greater than ten times its virial radius away, $D_{1,1} > 10$, and clustered
otherwise. The boundary between the low- and high-mass bins is at $2\times 10^{12}\, h^{-1}M_\odot$. As before, the results are presented for two halo separations, $r=1$ and $5\, h^{-1}$Mpc. We note that in some cases there are too few halo pairs at $1\, h^{-1}$Mpc separation in the isolated bin, so these bins were removed from the plot.

\begin{figure*}[t]
\centering
 \includegraphics[width=0.75\textwidth]{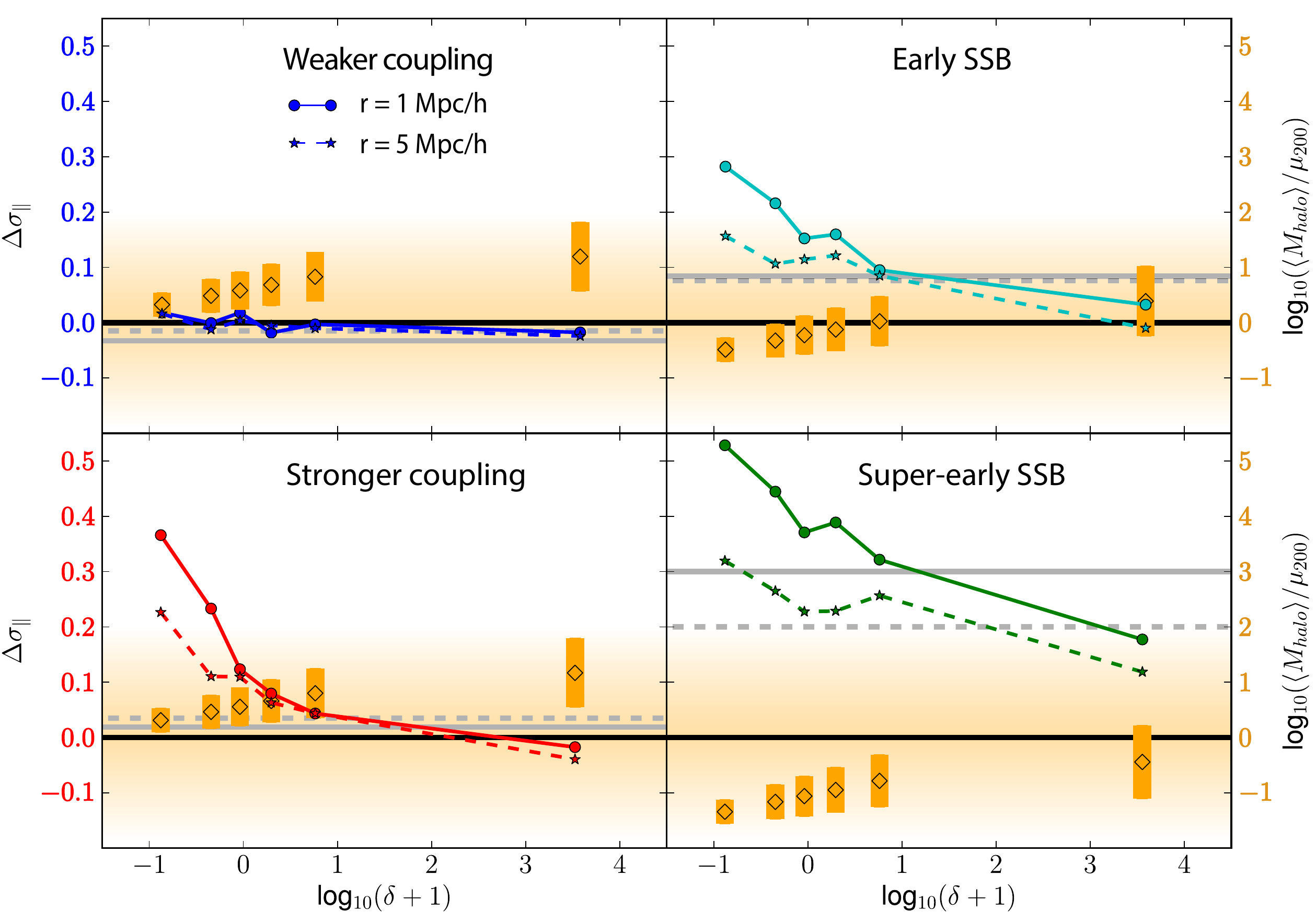}
  \caption{Fractional deviation from GR ($\Delta \sigma_\parallel = \sigma_{\parallel,\text{MG}}/\sigma_{\parallel,\text{GR}}-1$) for the four symmetron models (colored lines) for separations $r=1$ (solid lines) and 5 $h^{-1}$Mpc (dashed) as a function of ambient density, $\delta$. Deviations without binning in $\delta$ are shown as solid ($r~=~1h^{-1}$Mpc) and dashed ($r~=~5h^{-1}$Mpc) horizontal gray lines. Yellow points are the ratio of the mean halo mass to the rescaling mass (right axis; see Sect.~\ref{rescmass}), with the bars denoting the standard deviation within the bin, and the orange shading representing the band of partial screening found in \cite{gronke_universal_2015}.
}
\label{densfig}
\end{figure*}

Figure~\ref{envirfig} shows $\Delta\sigma_\parallel$ for the $f(R)$ models. For separations of $r=5\,h^{-1}$Mpc, $\Delta\sigma_\parallel$ depends strongly on $f_{R0}$ for both clustered and isolated halos; the weak $f(R)$ model ($f_{R0}\!=\!10^{-6}$) hardly deviates from GR, while strong $f(R)$ ($f_{R0}\!=\!10^{-4}$) shows a $\sim\! 30\%$ deviation in both mass bins. At this separation, we see no clear dependence of $\Delta \sigma_\parallel$ on isolatedness, although there is tentative evidence for mass-dependence in the isolated bin (the strong $f(R)$ model gives $\Delta \sigma_\parallel \approx 35\%$ and $23\%$ in the low- and high-mass bins, respectively). This can potentially be explained by noise, however, as the isolated bins contain relatively fewer halo pairs. Noise also explains the relative scatter in the results for $r=1\,h^{-1}$Mpc.

A clear dependence on isolatedness is found for one of the symmetron models, however (Fig.~\ref{envirfig2}). For $r=5\,h^{-1}$Mpc, the deviation for the stronger coupling symmetron model in the isolated bin is significant ($\sim\!\! 10-15\%$), but goes to zero in the clustered bin. A mild dependence on mass can also be seen in the isolated bin. The effect is much smaller for the weaker coupling model, which goes from a slightly positive (isolated) to slightly negative (clustered) deviation. These two models have the same $z_{\rm SSB}=1$ but differ in coupling strength by a factor of 2. For the other two symmetron models, there is no strong dependence on isolatedness; early SSB shows only a slight decrease in $\Delta \sigma_\parallel$ from the isolated to clustered bin, while a slight increase is observed for super-early SSB. The super-early SSB model shows a higher deviation overall.

Since SSB of the background occurred at a much earlier stage in these models ($z_{\rm SSB} = 2$ and $3$ ), we can expect the clustering of dark matter halos to have evolved with an active coupling between the scalar field and matter for a long time regardless of environment on cosmological scales, with screening only having been significant on very small scales. The environment
dependence seen in the stronger coupling model may therefore have been erased \citep{gronke_gravitational_2014}.

\begin{figure}[h]
\centering
\includegraphics[width=0.5\textwidth]{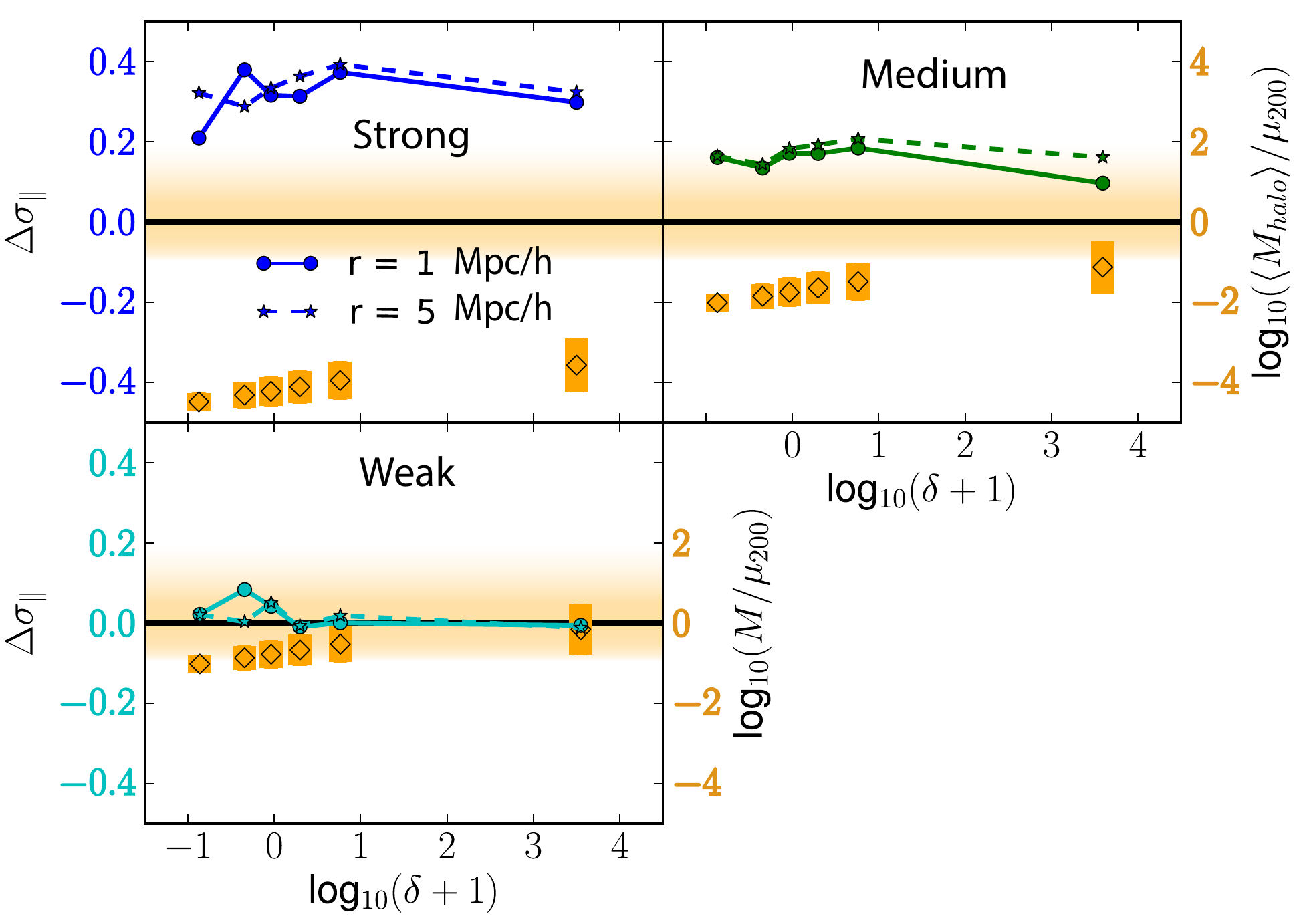}
\caption{Fractional deviation $\Delta\sigma_{\parallel}$ for the three $f(R)$ models (colored lines; left axis), and the ratio of mean halo mass to rescaling mass (yellow points; right axis); see Fig.~\ref{densfig} for key. 
}
\label{densfofr}
\end{figure}

\subsection{Dependence on ambient density} \label{sec:ambient}

Figures~\ref{densfig} and \ref{densfofr} show the fractional deviation $\Delta \sigma_\parallel$ for each theory as a function of ambient density contrast, $\delta$ (see Sect.~\ref{sec:density}). Also plotted is the mean mass of halos in each bin, $\langle M_h \rangle$, divided by the rescaling mass $\mu_{200}$ for each theory (see Sect.~\ref{rescmass}). Bins with $\hat{\mu} = \log_{10}(\langle M_h\rangle / \mu_{200}) \gg 0$ are expected to contain halos that are screened, while highly negative values of $\hat{\mu}$ are in the unscreened regime and therefore should correlate with stronger modified gravity effects \citep{gronke_universal_2015}. Between these two extremes, a band around $\hat{\mu}=0$  corresponds to partial screening (orange shading in the figures). In this region, the screening mechanism is active, but deviations from GR are not fully suppressed; this is the \textup{{\it \textup{transition region}}} between the fully screened and unscreened states.

This relationship between screening and rescaled halo mass $\hat{\mu}$ appears to be borne out by the figures. The $\Delta\sigma_\parallel$ curves for the strong and medium $f(R)$ theories (Fig.~\ref{densfofr}) are flat, indicating no significant dependence on $\delta$. Since $\hat{\mu}$ is below the screening transition region for almost all of the ambient density bins, this is to be expected: both of these theories are fully unscreened across the whole range of $\delta$, which means that no environment-dependent screening is taking place. This is not the case for the weak $f(R)$ model, for which most of the bins are inside the transition region, screening should be partially active, but this also exhibits a flat $\Delta\sigma_\parallel$ curve. This model gives too small a modification in the first place for any environment dependence to be noticeable, however.

The symmetron models show a broader range of behavior (Fig.~\ref{densfig}). In all cases and for all ambient density bins, $\hat{\mu}$ is within the screening transition region (which is also wider than for the $f(R)$ models). This suggests that environment-dependent screening should be observable for all four models, and this is indeed the case: the stronger coupling, early SSB, and super-early SSB models all show a strong evolution of $\Delta\sigma_\parallel$ with ambient density. The largest deviations occur in low-density (void) regions, with differences from GR+$\Lambda$CDM of up to 50\% being observed for the super-early SSB model. The deviations are much smaller at high values of $\delta$, as the symmetry of the scalar field potential is restored and the fifth force is suppressed. A $\sim\!\!20\%$ deviation is observed even in the $\delta \ge 10$ bin for super-early SSB, however, as this model is generally closer to being unscreened than the others. The weaker coupling symmetron model does not fit into this pattern, showing $\Delta\sigma_\parallel \approx 0$ across the full $\delta$ range. As with the weak $f(R)$ model, this may have more to do with the relative weakness of the modification to GR in this theory (although we note that it is not a weak modification in an absolute sense, as the matter power spectrum deviates from its GR behavior significantly on non-linear scales).

While the value of $\hat{\mu}$ relative to the screening transition region appears to give a satisfactory explanation of our results, we cannot exclude the possibility that the behavior of the $\Delta \sigma_\parallel$ curves is unrelated to the mass of the individual halos. An alternative explanation is that (a) $\Delta \sigma_\parallel$ responds directly to the screening state in the region between{\it } the halos, and (b) the link between $\Delta \sigma_\parallel$ and halo mass is a secondary effect caused by the weak dependence of halo mass on ambient density. In this scenario, the internal structure or screening state of the halos themselves does not matter; the mass dependence is just a coincidence which is due to the expectation from the peak-background split formalism \citep{1986ApJ...304...15B} that high-mass halos can form more readily in higher-density environments (see Table~\ref{bins} for the mean halo mass and bias of each ambient density bin).

If this were the case, splitting each ambient density bin into high- and low-mass samples would yield similar $\Delta \sigma_\parallel$ curves. The dependence of $\sigma_\parallel$ on mass is more or less flat (Sect.~\ref{sec:mass}), which means that if ambient density is the only factor, the results for the two mass bins should be the same. Conversely, if the value of $\hat{\mu}$ with respect to the transition regime is important, then the different mass bins will give different results depending on how deep they are into the transition region.

Regardless of the reason, Figs.~\ref{densfig} and \ref{densfofr} show a clear environmental dependence of $\Delta \sigma_\parallel$ that differs between modified gravity theories. This is a tell-tale signature of screening at work. Moreover, all of the quantities involved are observables, making it possible to look for these signatures in forthcoming velocity surveys.

\section{Discussion} \label{conclusions}

We used N-body simulations of seven different modified gravity theories (three $f(R)$ and four symmetron models) to investigate the dependence of a pairwise peculiar velocity statistic (the relative radial velocity dispersion, $\sigma_\parallel$) on halo environment. Strong environmental dependence is a key property of {\it \textup{screening mechanisms}} -- non-linear effects that are invoked by many MG theories to evade stringent local constraints on deviations from GR. By binning $\sigma_\parallel$ by a measure of environment  (ambient density),  we were able to observe signs of a transition from the screened to unscreened regimes for several of the theories we considered. The signatures differ markedly between models, with the $f(R)$ theories presenting an essentially constant deviation of $\sigma_\parallel$ from its GR value, and the symmetron models showing a strong evolution of the deviation with density.

We chose to focus on the $\sigma_\parallel$ statistic because it is sensitive to modified-gravity effects on the typical scale of the screening transition, that is, around the inter-halo separation scale. A related statistic, the line-of-sight velocity dispersion $\sigma_{\rm los}$, was also shown to be sensitive to modified gravity effects in \cite{hellwing_clear_2014}, although its environmental dependence was not considered. It is also less directly related to the relative radial velocity, $v_\parallel$, a physically meaningful quantity on which many velocity statistics are based. We compared these statistics in Sect.~\ref{velstat}, where we also described an estimator for $\sigma_\parallel$ that makes this quantity observable.

In Sect.~\ref{envircompar} we showed that two measures of halo environment, the isolatedness of halos, $D_{1,1}$, and the ambient density contrast, $\delta$, are fundamentally different, contrary to some claims in the literature. To study screening effects at least, we obtained more coherent results when binning by ambient density (Sect.~\ref{sec:ambient}). This matches expectations for symmetron theories, where the local density is explicitly responsible for restoring the broken symmetry that activates the screening mechanism (Eq.~\ref{eq:Vsymm}). Binning in isolatedness also revealed an environmental dependence on $\sigma_\parallel$, but the results were harder to interpret.

Throughout, we have worked under the assumption that noise-free distances and velocities are available for all dark matter halos down to some mass cutoff. Typical distance or velocity errors from peculiar velocity surveys are of the order of tens of percent, however \citep{2013AJ....146...86T}, and scale poorly with redshift, which limits the survey volume. As such, current precision peculiar velocity samples are relatively limited in size, with the largest compilations containing only $\sim {\rm \text{ a }   f ew} \times 10^3$ objects \citep{2013AJ....146...86T}, which  is most likely too small to see these effects. In addition, real-world measurements are obviously performed on galaxies and not dark matter halos, which complicates the process of associating subsamples with distinct mass bins. This also introduces a selection function that must be taken into account when (e.g.) correcting for Malmquist-type biases \citep{ferreira_streaming_1999}. Selection effects will presumably also be important in the reconstruction of the ambient density field (Sect.~\ref{sec:density}). Finally, the simulations we used did not take into account baryonic effects, which could also introduce environment-dependent signatures into the velocity field {\corr (although the recent work by \citet{2016arXiv160303328H} suggests that baryonic effects have a relatively small effect on halo velocities at the scales considered here).}

While an observational campaign to detect environment-dependent screening effects in the velocity field would clearly be more complex than the simple analysis we have presented here, there are no fundamental obstacles to its execution. All of the quantities we considered are closely related to observables {\corr (cf. Fig.~\ref{fig:sigparest})} and can be accessed through the design of appropriate estimators {\corr (Appendix~\ref{app:estimators})}. Furthermore, large homogeneously selected peculiar velocity surveys are planned for the near future, with optical and radio surveys expected to yield samples of $1-5 \times 10^4$ objects over a substantial fraction of the sky \citep{2014MNRAS.445.4267K}. Alternative methods of measuring the velocity dispersion in an environment-dependent manner, for example, with redshift-space distortions, may also prove viable. {\corr We leave the question} of the practical observability of these effects to future work.

In conclusion, we have presented a possible method for observing the screening transition in modified-gravity theories using the relative velocities of pairs of halos or galaxies and a density-based measure of halo environment. The signatures of deviations from GR are clear and depend on the particular modified gravity theory. The details of the signal and its ability to distinguish between different signals, will also depend on the properties of the peculiar velocity survey that is used, however; an investgation of this dependence is left to future work.

\FloatBarrier

\noindent {\it Acknowledgements:} We are grateful to A.~Barreira, B.~Falck, M.~Gr\"onke, {\corr P.~Lilje, H.~Winther, H.-Y.~Wu, and especially the referee, W.~Hellwing,} for useful comments and discussions, and R.O.~Fauli for preliminary work during a masters project. PB's research was supported by an appointment to the NASA Postdoctoral Program at the Jet Propulsion Laboratory, California Institute of Technology, administered by Universities Space Research Association under contract with NASA. DFM is supported by the Research Council of Norway. We thank NOTUR for proving computational support. 

\appendix

\section{Pairwise velocity estimators} \label{app:estimators}

For a given halo pair geometry, the variance of the relative velocity of the pair, $\vec{v}_i - \vec{v}_j$, projected along the LOS, is
\begin{equation}
\sigma_{\rm los}^2 = \frac{1}{2}\sin^2\theta\,\tilde{\sigma}_\perp^2(r, l) + \cos^2\theta\,[\tilde{\sigma}_\parallel^2(r,l) - (\tilde{v}_\parallel(r,l))^2],
\end{equation}
where we have used $\langle v_\perp \rangle = 0$. An empirical estimate of this quantity can be obtained by summing over many halo pairs with the same geometry,
\begin{equation}
\sigma_{\rm los, obs}^2 = \sum_{i,j} (s_i - s_j)^2 - \left ( \sum_{i,j} (s_i - s_j) \right )^2.
\end{equation}
Following a similar strategy to \cite{ferreira_streaming_1999}, we construct a simple estimator for $\sigma^2_\parallel$ by writing $\chi^2 = \sum ( \sigma_{\rm los, obs}^2 - \sigma_{\rm los}^2)^2$, where the sums are now over all pairs with fixed separation, $r$. We then minimize with respect to $\sigma^2_\parallel$ by setting $d\chi^2/d(\sigma^2_\parallel) = 0$, and rearrange to obtain
\begin{equation} \label{eq:sigpar_est}
\tilde{\sigma}^2_\parallel = \frac{\sum_{i,j} \cos^2 \theta_{ij} \left [ (s_i - s_j)^2 - \frac{1}{2}\tilde{\sigma}^2_\perp \sin^2 \theta_{ij} - \tilde{v}_\parallel^2 (1 - \cos^2\theta_{ij}) \right ]}{\sum_{i,j} \cos^4\theta_{ij}}.
\end{equation}
An estimate of $\tilde{v}_\parallel$ can be obtained using Eq.~(\ref{v12-est}), and  $\tilde{\sigma}^2_\perp$ by performing the minimization $d\chi^2/d(\sigma^2_\perp) = 0$. After rearranging and combining with Eq.~(\ref{eq:sigpar_est}), we obtain
\begin{multline}
\tilde{\sigma}^2_\parallel(r) = \left[ \left(\sum_{i,j} \sin^4 \theta_{ij}\right) \left(\sum_{i,j} \cos^2 \theta_{ij} (s_i - s_j)^2\right) - \right. \\   \left. \left(\sum_{i,j} \cos^2 \theta_{ij} \sin^2 \theta_{ij}\right)  \left(\sum_{i,j} \sin^2 \theta_{ij} (s_i - s_j)^2\right) \right] \times \\ \left[\left(\sum_{i,j} \cos^4 \theta_{ij}\right) \left(\sum_{i,j} \sin^4 \theta_{ij}\right) - \left(\sum_{i,j} \cos^2 \theta_{ij} \sin^2 \theta_{ij}\right)^2 \right]^{-1}
\end{multline}
and
\begin{multline}
\tilde{\sigma}^2_\perp(r) = 2\, \left[ \left(\sum_{i,j} \cos^4 \theta_{ij}\right) \left(\sum_{i,j} \sin^2 \theta_{ij} (s_i - s_j)^2\right) - \right. \\
\left. \left(\sum_{i,j} \cos^2 \theta_{ij} \sin^2 \theta_{ij}\right) \left(\sum_{i,j} \cos^2 \theta_{ij} (s_i - s_j)^2\right) \right]\times \\
\left[  \left(\sum_{i,j} \cos^4 \theta_{ij}\right) 
 \left(\sum_{i,j} \sin^4 \theta_{ij}\right) - \left(\sum_{i,j} \cos^2 \theta_{ij} \sin^2 \theta_{ij}\right)^2 \right]^{-1} - 2\,\tilde{v}_\parallel^2(r).
\end{multline}
Note that these expressions do not account for the noise in the LOS velocity measurements, which will increase the variance (and therefore must be taken into account in any real measurement).

\FloatBarrier

\balance

\bibliographystyle{aa}
\bibliography{draft-aa_v1}

\end{document}